\documentclass[journal]{IEEEtran}

\ifCLASSINFOpdf
\else
   \usepackage[dvips]{graphicx}
\fi
\usepackage{url}
\usepackage{amssymb,amsmath,bm}
\usepackage{textcomp}
\usepackage{booktabs}

\usepackage{graphicx}

\begin{document}

\title{Zero-Shot Voice Conversion via Content-Aware Timbre Ensemble and Conditional Flow Matching}

\author{Yu Pan, Yuguang Yang, Jixun Yao, Lei Ma, Jianjun Zhao
\thanks{This work has been submitted to the IEEE for possible publication. Copyright may be transferred without notice, after which this version may no longer be accessible.}
}

\markboth{Journal of \LaTeX\ Class Files, Vol. 14, No. 8, August 2015}
{Shell \MakeLowercase{\textit{et al.}}: Bare Demo of IEEEtran.cls for IEEE Journals}
\maketitle

\begin{abstract}
Despite recent advances in zero-shot voice conversion (VC), achieving speaker similarity and naturalness comparable to ground-truth recordings remains a significant challenge.
In this letter, we propose \textbf{\textit{CTEFM-VC}}, a zero-shot VC framework that integrates \textbf{C}ontent-aware \textbf{T}imbre \textbf{E}nsemble modeling with conditional \textbf{F}low \textbf{M}atching.
Specifically, CTEFM-VC decouples utterances into content and timbre representations and leverages a conditional flow matching model to reconstruct the Mel-spectrogram of the source speech. 
To enhance its timbre modeling capability and naturalness of generated speech, we first introduce a context-aware timbre ensemble modeling approach that adaptively integrates diverse speaker verification embeddings and enables the effective utilization of source content and target timbre elements through a cross-attention module. 
Furthermore, a structural similarity-based timbre loss is presented to jointly train CTEFM-VC end-to-end.
Experiments show that CTEFM-VC consistently achieves the best performance in all metrics assessing speaker similarity, speech naturalness, and intelligibility, significantly outperforming state-of-the-art zero-shot VC systems.
\end{abstract}


\IEEEpeerreviewmaketitle

\section{Introduction}

\IEEEPARstart{A}{s} a pivotal task within the field of speech signal processing, zero-shot voice conversion (VC) aims to transfer the timbre of a source utterance to an arbitrary unseen speaker while maintaining the original phonetic content, with applications spanning various practical domains such as voice anonymization \cite{yao2024musa} and audiobook production \cite{chen2024takin}.

Generally, the primary difficulties of zero-shot VC lie in effectively modeling, decoupling, and utilizing various speech attributes, including content, timbre, etc.
Previous methods \cite{tan2021zero,zhao2022disentangling,kovela2023any} often use pre-trained automatic speech recognition (ASR) methods \cite{gulati2020conformer,kim2022squeezeformer} and speaker verification (SV) models \cite{desplanques2020ecapa,wang2023cam++} to extract linguistic content and timbre information from the source and target speech, respectively. 
Nevertheless, due to many factors such as the inherent complexity of speech signals \cite{pan2024gemo,pan2024gmp} and limitations in timbre and content modeling methods \cite{hussain2023ace,li2024sef}, they still remain significant potential for performance enhancement.
With the progressions of self-supervised learning (SSL)-based speech models \cite{hsu2021hubert,chen2022wavlm}, many works \cite{dang2022training,hussain2023ace} have sought to use them to extract semantic features from speech. However, these extracted features inevitably contain certain characteristics of the source timbre, which ultimately affect their VC quality. 
To this end, \cite{yao2024promptvc,li2024sef,yang2024takin} incorporated the model quantization technique to minimize noncontent elements. 
Nonetheless, this operation may lead to token format inconsistencies between different models, thus impairing the overall performance and limiting the broader applicability.
Moreover, existing zero-shot VC methods \cite{tan2021zero,kovela2023any,hussain2023ace} normally employ a single pre-trained SV model to capture target timbre embeddings. Although speaker embedding techniques have advanced significantly \cite{wang2023cam++,chen2023enhanced,yakovlev2024reshape}, relying exclusively on a single model is insufficient to deliver optimal VC performance, resulting in sub-par speaker similarity compared to authentic recordings.

Inspired by the powerful zero-shot capabilities of recent large-scale language models (LLMs) \cite{liu2023gpt,bai2023qwen}, several studies \cite{wang2023lm,wang2024streamvoice} have attempted to discretize waveforms using neural codecs \cite{defossez2022high,pan2024promptcodec} and then leverage LLMs to generate target waveforms in an autoregressive manner. \cite{wang2023lm} proposed a two-stage language model that first generates coarse acoustic tokens to capture the source content and target timbre elements, and then refines the acoustic details for VC. \cite{wang2024streamvoice} implemented a single stage VC framework based on the context-sensitive language model and acoustic predictor, which facilitated zero-shot VC in a streamable way. 
Despite great results, such methods commonly encounter stability problems due to their auto-regressive fashion and may experience error accumulation, leading to a gradual decline in VC performance.

To address the aforementioned problems, we propose \textbf{\textit{CTEFM-VC}}, a novel zero-shot VC framework based on content-aware timbre ensemble modeling and conditional flow matching.
Specifically, we first present a simple yet effective timbre ensemble modeling approach that adaptively integrates multiple pre-trained SV models to robustly capture speaker-specific characteristics from the reference speech. Subsequently, we introduce a content-aware timbre ensemble (CTE) method that exploits linguistic content features extracted from a pre-trained ASR model \cite{yang2023hybridformer} to dynamically fuse timbre representations via an advanced cross-attention module, thereby enhancing both speech naturalness and speaker similarity. To further improve the naturalness and enable stable training, we develop a conditional flow matching model and adopt a pre-trained vocoder \cite{lee2022bigvgan} to reconstruct the mel-spectrogram features of the source speech and generate the converted waveform, respectively. Finally, to further boost speaker similarity, we propose a structural similarity (SSIM)-based timbre loss for joint optimization of the overall system.
\begin{figure*}[htbp]
\centering
    \includegraphics[height=7.0cm,width=!]{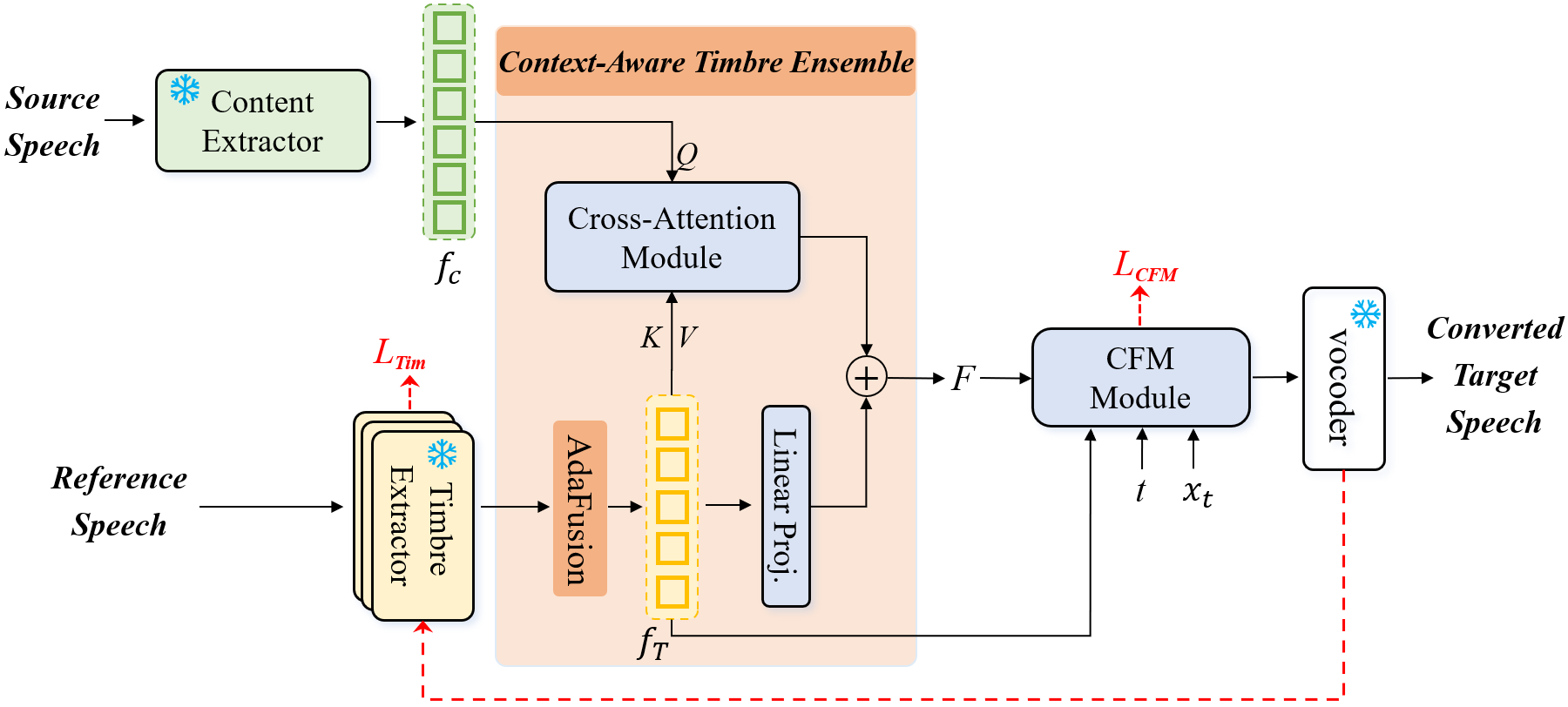}
    \caption{Overall training architecture of the proposed CTEFM-VC. $Q$, $K$, $V$ denote the query, key, and value of the cross-attention module, respectively.}
    \label{fig:ctefmvc}
\end{figure*}
Extensive experiments demonstrate that CTEFM-VC surpasses several state-of-the-art (SOTA) zero-shot VC systems in terms of speaker similarity, speech naturalness, and intelligibility.

\section{METHODOLOGY}
\label{sec:METHODOLOGY}
\subsection{System Architecture}

As depicted in Fig. \ref{fig:ctefmvc}, the proposed CTEFM-VC is an end-to-end zero-shot VC framework.
Assume the input speech signal is represented as $X = [x_1, x_2, ..., x_T]$ $\in$ $R^T$, our CTEFM-VC initially adopts a pre-trained ASR method, HybridFormer \cite{yang2023hybridformer}, to extract the linguistic content $f_C$ $\in$ $R^{T_1 \times D}$.
Detailed, the used HybridFormer consists of 12 blocks, each with a convolution kernel size of 31 and 4 attention heads. The hidden dimensions of the attention layer and feedforward network (FFN) are set to 256 and 1024, respectively.
For timbre modeling, multiple pre-trained SV models are employed \cite{wang2023cam++,chen2023enhanced,yakovlev2024reshape} to extract their corresponding timbre embeddings $f_{T_i}$ $\in$ $R^{d_i}$ from the reference utterance.
To further enhance speaker similarity and speech naturalness, 
we propose CTE, a context-aware timbre ensemble modeling approach that uses a straightforward yet effective AdaFusion method to fuse all SV embeddings $f_{T_i}$ as global representations $f_T$ and introduce an advanced cross-attention module to facilitate joint utilization of $f_C$ and $f_T$.
Last, the outputs $F$ of CTE and the ensembled target timbre characteristics $f_T$ are fed into a conditional flow matching model to reconstruct the Mel-spectrogram of source speech, followed by a pre-trained Bigvgan vocoder to generate the converted target waveform.

\subsection{CTE: Context-aware Timbre Ensemble Modeling}
Essentially, the core challenge of zero-shot VC stems from timbre modeling, as it requires the model to generalize effectively to any previously unseen speakers without additional training or fine-tuning. 
To address this, we first propose a model ensemble strategy that integrates multiple pre-trained SV models to extract timbre features. We hypothesize that 
\begin{figure}[h]
\centering
	\includegraphics[height=9cm]{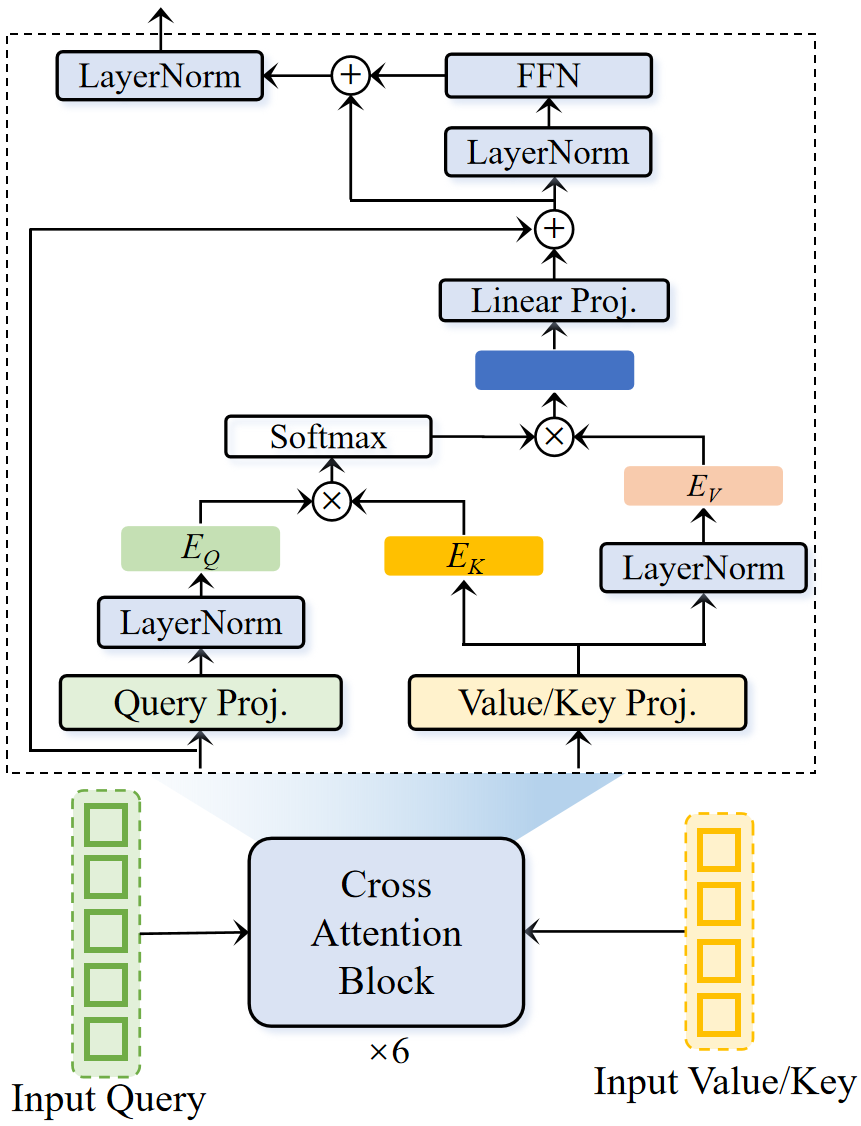}
	\caption{\label{fig:ca} Schematic of the proposed cross-attention module.}
\end{figure}
integrating diverse timbre embeddings enables our approach to capture a richer array of timbre characteristics, thereby enhancing its adaptability to variations in speaker identity.

In addition, to better exploit these feature representations, we propose a simple yet effective AdaFusion approach that applies learnable hyperparameters to weight these SV embeddings before concatenating them into a unified global timbre representation $f_T$.

Subsequently, to facilitate the adaptive utilization of the captured source linguistic content and target timbre characteristics, we propose a cross-attention module comprising six multi-head cross-attention blocks, as depicted in Fig. \ref{fig:ca}. Each block consists of a stack of linear projection layer, layer normalization layer, and FFN, where $f_C$ serves as the query in the cross-attention mechanism and $f_T$ functions as the key and value. 

Ultimately, the $f_T$ is projected to match the dimensionality of the cross-attention module's output via a linear layer and then added element-wise to the output features of the cross-attention module, as shown in Fig.~\ref{fig:ctefmvc}.

\subsection{Optimal Transport-based Conditional Flow Matching}

To improve the speech naturalness and ensure stable training, we incorporate an optimal transport (OT)-based CFM module to reconstruct the target mel-spectrogram $x_1 \!= p_1(x)$ from standard Gaussian noise $x_0 = p_0(x) \!= \mathcal{N}(x;0,I)$. 
Concretely, an OT flow ${\psi}_{t}: [0, 1] \times R^d \rightarrow R^d$ is used to train our proposed OT-CFM model that is composed of multiple UNet \cite{ronneberger2015u} blocks with timestep fusion.
By adopting the ordinary differential equation to estimate a learnable time-dependent vector field $v_t\!: [0, 1] \times R^d \rightarrow \!R^d$, it can approximate the optimal transport probability path from $p_0(x)$ to the target distribution $p_1(x)$:
\begin{equation}
    \label{eq:ode}
    \begin{split}
        \frac{d}{d_t} \psi_t(x)=v_t(\psi_t(x),t)
    \end{split}
\end{equation}
where $\psi_0(x)=x$, and $t \in [0, 1]$. 
Additionally, inspired by \cite{tong2023conditional} that recommends straighter trajectories, we simplify the formula of the OT flow as follows: 
\begin{equation}
    \label{eq:111}
    \begin{split}
        \psi_{t,z}(x) = \mu_t(z) + \sigma_t(z)x \quad \quad \quad \\ 
    \end{split}
\end{equation}
where $\mu_t(z) = tz$, $\sigma_t(z) = (1 \!-\! (1\!-\!\sigma_{min})t)$, $z$ represents the random conditioned input, $\sigma_{min}$ signifies the minimum standard deviation of white noise introduced to perturb individual samples, empirically set to 0.0001. 
As a result, the training loss of our OT-CFM module is formulated as:
\begin{equation}
    \mathcal{L}_{C\!F\!M} \!= \mathbb{E}_{t,p(x_0),q(x_1)}\Vert(x_1\!-(1\!-\sigma)x_0)\!-v_t({\psi_{t,x_1}}(x_0)|h)\Vert^2
\end{equation}

Here, $t \sim U[0,1]$, $x_0 \sim p(x_0)$, $x_1 \sim q(x_1)$, with $q(x_1)$ representing the true but potentially non-Gaussian distribution of the data, and $h$ denotes the captured conditional inputs of the CFM model.

\subsection{Training Objectives}

In addition to the $L_{C\!F\!M}$, we incorporate a structural similarity-based loss function to minimize the disparity between the timbre embeddings of the reference and converted speech, thus further enhancing the speaker similarity performance of CTEFM-VC:
\begin{equation}
    \begin{gathered}
    L_{Tim} = \sum_{i=1}^N L_{Tim_i} = -\sum_{i=1}^N \text{SSIM}(T_{r_i}, T_{c_i}) \\
    \text{SSIM}(T_{r_i}, T_{c_i}) \!= \frac{(2\mu_{T_{r_i}} \mu_{T_{c_i}} \!+\! c_1)(2\sigma_{{T_{r_i}}{T_{c_i}}} \!+\! c_2)}{(\mu_{T_{r_i}}^2 \!+\! \mu_{T_{c_i}}^2 \!+\! c_1)(\sigma_{T_{r_i}}^2 \!+\! \sigma_{T_{c_i}}^2 \!+\! c_2)}
    \end{gathered}
\end{equation}
where N is 3, $T_{r_i}$ and $T_{c_i}$ are the corresponding features of the SV model, $\mu_{T_{r_i}}$ and $\mu_{T_{c_i}}$ denote the means of $T_{r_i}$ and $T_{c_i}$, while $\sigma_{T_{r_i}}^2$ and $\sigma_{T_{c_i}}^2$ represent their respective variances.
In addition, $\sigma_{{T_{r_i}}{T_{c_i}}}$ indicates the covariance between $T_{r_i}$ and $T_{c_i}$. The constants $c_1$ and $c_2$ are used to ensure numerical stability during division, with values set to $0.01$ and $0.03$, respectively.

Therefore, the overall training objective of the proposed CTEFM-VC can be expressed as:
\begin{equation}
    \begin{split}
        L_{Total} = \mathcal{L}_{C\!F\!M} + \lambda L_{Tim}
    \end{split}
\end{equation}
\noindent
where $\lambda$ is a tuning hyperparameter empirically set to 0.05.

\section{EXPERIMENTS}
\label{sec:EXPERIMENTS}

\subsection{Experimental Setups}

\subsubsection{Datasets}
We conduct all experiments on the LibriTTS \cite{zen2019libritts} dataset, which comprises 585 hours of English recordings from 2,456 speakers. 
All data is downsampled to 16kHz. We use all training subsets for model training, while the dev-clean subset is employed for validation. 
To assess their zero-shot VC performance, we select the VCTK \cite{yamagishi2019cstr} and ESD \cite{zhou2022emotional} corpora, following the settings in \cite{yao2024stablevc}. For each corpus, we randomly select 100 samples from 10 unseen speakers, ensuring that there is no overlap with the training data.

\subsubsection{Implementation Details}

We adopt the AdamW optimizer to train the proposed CTEFM-VC approach over 600K iterations using four NVIDIA A10 GPUs, with an initial learning rate of 1e-4 and a batch size of 64. During training, we randomly segment 4s of the source speech as the reference waveform. During the inference stage, the CFM model operates with 20 Euler steps to generate the target outputs.

\subsubsection{Evaluation Metric}

To thoroughly assess our CTEFM-VC method, we perform objective and subjective evaluations. 

In the objective evaluation, we compute the speaker embedding cosine similarity (SECS), word error rate (WER), and UTMOS between the converted speech and the reference speech using a pre-trained WavLM-TDCNN model$^1$, a CTC-based ASR system$^2$, and a mean opinion score (MOS) prediction model$^3$, respectively.
Regarding subjective evaluation, we invite 15 professional raters to assign MOS scores for both naturalness (NMOS) and similarity (SMOS) on a scale ranging from 1 to 5. This scoring reflects the naturalness of the generated speech and the speaker similarity between the converted and source waveforms. A score of '5' indicates excellent quality, '4' denotes good quality, '3' represents fair quality, '2' indicates poor quality, and '1' signifies bad quality.
Audio samples are available online$^4$.

{
\let\thefootnote\relax
\footnote{$^1$https://github.com/microsoft/UniSpeech/tree/main/downstreams/\\speaker\_verification}
\footnote{$^2$https://huggingface.co/facebook/hubert-large-ls960-ft}
\footnote{$^3$https://github.com/tarepan/SpeechMOS}
\footnote{$^4$https://yupan0v0.github.io/ctefm-vc/}
}

\subsection{Main Results}

\begin{table*}[h]
\centering
\caption{Comparative results of subjective and objective evaluations between the proposed CTEFM-VC and baseline zero-shot \\ VC systems. Subjective metrics are computed with 95\% confidence intervals and 'GT' denotes the ground truth recordings. \\ The best results are highlighted in bold, while the sub-optimal results are underlined.}\label{tab:main}
\begin{tabular}{lcccccc}
\hline
         & NMOS ($\uparrow$)& SMOS ($\uparrow$) & WER ($\downarrow$) & UTMOS ($\uparrow$) & SECS ($\uparrow$) \\ \hline
GT       & 4.18$\pm0.05$ & -    & 2.01 & 4.19  & - &   \\ \hline
DiffVC \cite{popov2021diffvc}  & 3.75$\pm0.05$ & 3.66$\pm0.07$ & 3.08 & 3.68  & 0.61   \\
NS2VC  \cite{shen2023naturalspeech}  & 3.65$\pm0.07$ & 3.51$\pm0.06$ & 2.94 & 3.64  & 0.53   \\
VALLE-VC \cite{wang2023valle} & 3.80$\pm0.06$ & 3.79$\pm0.04$ & \underline{2.77} & 3.72  & 0.65   \\
SEFVC \cite{li2024sef}   & 3.68$\pm0.05$ & 3.76$\pm0.06$ & 3.75 & 3.51  & 0.63   \\
StableVC \cite{yao2024stablevc} & \underline{3.83}$\pm0.04$ & \underline{3.88}$\pm0.06$ & \underline{2.77} & \underline{3.92}  & \underline{0.66}    \\
CTEFM-VC & \textbf{3.92}$\pm0.05$ & \textbf{4.16$\pm0.04$} & \textbf{2.41} & \textbf{3.99}  & \textbf{0.78}   \\ \hline
\end{tabular}
\end{table*}

To examine the performance of the proposed CTEFM-VC, we compare it with five SOTA zero-shot VC baselines, with the results reported in Table \ref{tab:main}.

Our experimental results show that CTEFM-VC achieves superior performance in both subjective and objective evaluations. 
In terms of objective assessment, CTEFM-VC consistently outperforms all comparative systems in speaker similarity, speech naturalness and intelligibility. 
In detail, CTEFM-VC attains a SECS of 0.78, exceeding other SOTA methods by at least 18.2\%. This result showcases the powerful capabilities of the proposed method in accurately converting the target timbre. In addition, our method achieves the lowest WER and the best UTMOS scores, indicating that the converted speech by our CTEFM-VC exhibits superior intelligibility and speech naturalness.

Regarding subjective evaluation, the CTEFM-VC approach consistently exceeds comparative zero-shot VC baselines in all evaluation metrics as well. 
In particular, CTEFM-VC achieves the highest SMOS of 4.16, which is correlated with the SECS metric and further validates its robust capabilities in converting the target timbre. With respect to NMOS, our proposed approach also gets the best score of 3.92.
In general, these findings provide compelling corroboration for subjective results, underscoring the effectiveness and robustness of the proposed CTEFM-VC.

\subsection{Ablation Study}

To validate the contributions of each component of CTEFM-VC, ablation studies are conducted. All results are summarized in Table \ref{tab:ablation}.

\begin{table}[h]
\centering
\caption{Results on ablation studies. 'w/o' denotes removing the corresponding module. 'SV1/2/3' are the pre-trained CAM++, ERes2Net, and ReDimNet models. 'w/o CTE' denotes just using the SV1 embeddings and $f_C$ as the condition and inputs to CFM.}\label{tab:ablation}
\begin{tabular}{lcccc}
\hline
         & NMOS ($\uparrow$)& SMOS ($\uparrow$) & WER ($\downarrow$)  & SECS ($\uparrow$) \\ \hline
CTEFM-VC & \textbf{3.92$\pm0.05$} & \textbf{4.16$\pm0.04$} & \textbf{2.41} & \textbf{0.78} \\
 \quad w/o \textit{CTE} & 3.63$\pm0.04$ & 3.71$\pm0.03$ & 2.89   & 0.60 \\
 \quad w/o \textit{SV1} & 3.66$\pm0.05$ & 3.79$\pm0.05$ & 2.85   & 0.65 \\
 \quad w/o \textit{SV2} & 3.71$\pm0.04$ & 3.85$\pm0.05$ & 2.78   & 0.68 \\
 \quad w/o \textit{SV3} & 3.73$\pm0.05$ & 3.92$\pm0.04$ & 2.76  & 0.69 \\
 \quad w/o \textit{AdaFusion} & 3.78$\pm0.05$ & 3.94$\pm0.05$ & 2.68  & 0.71 \\
 \quad w/o \textit{$L_{Tim}$} & \textbf{3.92$\pm0.04$} & 3.78$\pm0.05$ & 2.42  & 0.63 \\
\hline
\end{tabular}
\end{table}

Initially, we assess the effectiveness of the CTE strategy. 
As indicated in Table \ref{tab:ablation}, 
without our CTE method, all metrics of CTEFM-VC show a significant decline, demonstrating its effectiveness.
Then, the omission of any pre-trained SV model also leads to varying performance decrease across all metrics, particularly evident in the SECS score. 
These phenomena support our hypothesis that the integration of diverse timbre embeddings can facilitate the capture of a wider range of timbre characteristics, thereby enhancing its timbre modeling capabilities.
Furthermore, the experimental results indicate that within the CTEFM-VC framework, CAM++ shows the most effective timbre modeling capability, followed closely by ERes2Net, while ReDimNet exhibits the least effectiveness.

Next, we examine the influence of removing the AdaFusion method.
From the table, we can easily observe that in the absence of AdaFusion, both subjective and objective metrics show varying degrees of decline. Specifically, subjective NMOS and SMOS scores decrease by 3.6\% and 5.3\%, while objective SECS and WER drop by 9.0\% and 11.2\%. This reduction proves the effectiveness of the AdaFusion in adaptively fusing different timbre embeddings. By learning the importance of individual SV model, AdaFusion facilitates the joint utilization of linguistic content and timbre features, thus enhancing overall performance of CTEFM-VC.

Finally, we investigate the impact of $L_{Tim}$.
As shown in the table above, without the $L_{Tim}$, the SECS and SMOS drop significantly, while NMOS exhibits a slight improvement and WER shows a comparable performance.
These results suggest that the proposed $L_{Tim}$ is crucial to capture speaker-specific characteristics. However, the results in NMOS and WER without $L_{Tim}$ indicate that it could slightly enhance naturalness and intelligibility at the cost of speaker similarity. This trade-off highlights the importance of carefully balancing the loss components to achieve optimal performance.

\section{CONCLUSIONS}
\label{sec:CONCLUSIONS}
In this work, we propose CTEFM-VC, an innovative and scalable zero-shot VC workflow based on content-aware timbre ensemble modeling and conditional flow matching.  
To elaborate, CTEFM-VC utilizes a pre-trained ASR model and multiple SV models to extract linguistic content and timbre features. 
Subsequently, a content-aware timbre ensemble modeling method is proposed to integrate diverse timbre embeddings and facilitate adaptive utilization of the source content and target timbre. To enhance the speech naturalness and enable stable training, we incorporate a CFM model to reconstruct the source Mel-spectrogram, followed by a pre-trained vocoder to generate the converted speech. 
The entire system is jointly trained by the CFM loss and the SSIM-based timbre loss end-to-end. 
Extensive experiments on the LibriTTS corpus indicate that compared to recent SOTA zero-shot VC methods, CTEFM-VC consistently realizes superior speaker similarity, speech naturalness, and intelligibility.
Abaltion studies prove the effectiveness of each component.

\bibliographystyle{IEEEtran}
\bibliography{main}

\end{document}